\documentclass{article}
\usepackage[utf8]{inputenc}

\usepackage{authblk}

\title{Witnessing the Effective Entanglement in the COW Protocol}

\author[1]{F. Rezazadeh}
\author[2]{A. Mani}
\author[1]{M. Khodabandeh}
\author[1]{M. Jaberi}
\author[1]{S. A . Madani}

\affil[1]{Quantum Communication Group, Iranian Center for Quantum Technologies (ICQTs), Tehran, Iran}

\affil[2]{Department of Engineering Science, College of Engineering, University of Tehran, Iran}

\date{}

\usepackage{cite}
\usepackage{amsmath,amsfonts}
\usepackage{algorithmic}
\usepackage{algorithm}
\usepackage{array}
\usepackage[caption=false,font=normalsize,labelfont=sf,textfont=sf]{subfig}
\usepackage{graphicx}
\usepackage{stfloats}
\usepackage{url}
\usepackage{verbatim}
\usepackage{textcomp}
\usepackage{multirow}
\usepackage{float}
\usepackage{bm}
\usepackage{soul}
\usepackage{amsmath}
\usepackage{array}
\usepackage[table]{xcolor}   
\usepackage{booktabs}

\makeatletter
\AtBeginDocument{\DeclareMathVersion{bold}
\SetSymbolFont{operators}{bold}{T1}{times}{b}{n}
\SetMathAlphabet{\mathrm}{bold}{T1}{times}{b}{n}
\SetMathAlphabet{\mathit}{bold}{T1}{times}{b}{it}
\SetMathAlphabet{\mathbf}{bold}{T1}{times}{b}{n}
\SetMathAlphabet{\mathtt}{bold}{OT1}{pcr}{b}{n}
\SetSymbolFont{symbols}{bold}{OMS}{cmsy}{b}{n}
\renewcommand\boldmath{\@nomath\boldmath\mathversion{bold}}}
\makeatother

\def\BibTeX{{\rm B\kern-.05em{\sc i\kern-.025em b}\kern-.08em
    T\kern-.1667em\lower.7ex\hbox{E}\kern-.125emX}}

\begin{document}

\maketitle

\begin{abstract}

We present a rigorous mathematical framework for verifying effective entanglement in a Coherent One-Way (COW) quantum key distribution setup. In particular, we introduce a two-parameter family of entanglement witnesses, identify the parameter ranges where they constitute valid witnesses, and demonstrate their ability to reveal effective entanglement in the COW protocol. Additionally, we analyze previously obtained experimental data from a COW implementation and report clear signatures of effective entanglement.

\end{abstract}
	
\vskip 2cm

\section{Introduction}

Quantum Key Distribution (QKD) is one of the most prominent applications of quantum information science, enabling two distant parties to establish a secret cryptographic key with information-theoretic security. A typical QKD protocol consists of two main parts. The first is the quantum phase, during which quantum states are prepared, transmitted, and measured. The second is the classical post-processing, which involves parameter estimation, error correction, and privacy amplification — all carried out through classical communications.
\\
A key question of a QKD scenario is under what conditions the correlations obtained in the quantum phase can be turned into a secure key in the subsequent classical part. 
This question has been explored from multiple perspectives \cite{acin2005quantum,moroder2006upper,curty2004entanglement}.
For example, the authors of \cite{acin2005quantum} demonstrate a one-to-one correspondence between entanglement and secrecy, showing that entanglement is both necessary and sufficient for the generation of secret correlations. Building on this foundation, a practical method for estimating the maximum secret key that can be distilled from experimental QKD data is introduced in \cite{moroder2006upper}. Furthermore,  \cite{curty2004entanglement} establishes the necessary conditions for successful post-processing.  Specifically, for a typical QKD scenario where Alice measures the observable $A$, obtaining outcome $a$, and Bob measures the observable $B$, with outcome $b$, it is shown  that the joint probability distributions of the experiment, i. e. $P_{A,B}(a,b)$, cannot yield secret key bits unless it can be demonstrated that $P_{A,B}(a,b)$ originates from an (effective) entangled state.  \\  
 
This theorem is true for entanglement-based (EB) QKD protocols, as well as for prepare-and-measure (P\&M) protocols \cite{curty2004entanglement}. In EB scenarios, Alice and Bob share correlated states and perform measurements on their respective parts to obtain the probability distributions. Hence, security requires demonstrating that the shared state is indeed entangled—that is, the observed correlations $P_{A,B}(a,b)$ from the measurement results cannot be reproduced by any separable state.
In EB implementations, Alice and Bob share correlated quantum systems and obtain measurement results characterized by the joint distributions $P_{A,B}(a,b)$. Security requires proving that these correlations cannot arise from any separable state.
In P\&M protocols, no physical entangled state is distributed between Alice and Bob. Instead, Alice prepares and sends quantum states to Bob, who then performs measurements, and together they obtain probability distributions like $P_{A,B}(a,b)$. Such data can generate a secret key only if they can be regarded as originating from an entangled state - or equivalently, if the corresponding EB version of that P\&M protocol would involve an entangled shared state. This is referred to as \emph{effective entanglement} \cite{curty2005detecting, lorenz2006witnessing}.\\

Since experimental data provide only partial information about the nature of the QKD protocol or the underlying quantum state—rather than its full tomography—the verification of effective entanglement becomes a crucial task. In this context, entanglement witnesses \cite{horodecki1996necessary, terhal2000bell, lewenstein2000optimization} serve as powerful tools. These are operators whose expectation values are nonnegative for all separable states but take negative values for at least one entangled state.
The concept of entanglement witnessing has been explored in a variety of quantum key distribution (QKD) protocols. In \cite{curty2004entanglement}, the authors demonstrated that secure key generation during the post-processing phase necessitates the detectability of effective entanglement through an entanglement witness constructed from the measurement statistics of Alice and Bob. They further identified suitable witnesses for detecting entanglement in the two well-known entanglement-based protocols, namely BB84 and the six-state protocol. The framework of entanglement witnessing was subsequently extended to four-qubit and two-qubit prepare-and-measure protocols in \cite{curty2005detecting}. Related investigations have also been carried out for continuous-variable (CV) and qubit–mode hybrid protocols in \cite{grosshans2003virtual} and \cite{rigas2006entanglement}, respectively. In addition, an experimental observation of effective entanglement in a CV protocol was reported in \cite{lorenz2006witnessing}.\\

Although numerous theoretical and experimental studies have examined the manifestation of (effective) entanglement in various QKD scenarios, this question has not yet been addressed for the Coherent One-Way (COW) protocol which was first introduced in 2004 \cite{gisin2004towards,stucki2005fast,stucki2007coherent}. Notably, the COW protocol is among the most widely implemented QKD schemes, owing to its simplicity, cost-effective setup, practical advantages, and successful commercialization \cite{stucki2009continuous, eraerds2010quantum,stucki2009high,walenta2014fast,constantin2017fpga, roberts2017modulator, boaron2018secure, boaron2018simple, malpani2024implementation, shaw2022optimal, IDQ}.
In the present work, we investigate the witnessing of effective entanglement within the framework of the COW protocol. We first establish the theoretical formalism required to assess the presence of effective entanglement in experimental data obtained from a COW setup. Subsequently, we demonstrate that the probability distributions derived from our measurements (previously reported in \cite{dadahkhani2025experimental}) provide unambiguous evidence for the existence of effective entanglement.\\

The remainder of this paper is structured as follows. Section (\ref{COWsection}) provides a brief review of the COW protocol. In Section (\ref{theoretical}), we introduce a family of suitable witness operators for detecting entanglement within this protocol and derive the conditions under which they serve as valid entanglement witnesses. Section (\ref{experimental witnessing}) applies this theoretical framework to our experimental data, demonstrating the presence of effective entanglement in our setup. Finally, Section (\ref{conclosion}) concludes the paper with a discussion of the results and their implications.

\section{COW protocol}
\label{COWsection}

The COW protocol was originally designed to adapt quantum key distribution (QKD) to existing laboratory conditions. Instead of requiring a single-photon source, it relies on attenuated laser pulses, making it more practical for experimental implementation. \\

\begin{figure*}[!h]
	\centering
	\includegraphics[width=1\linewidth]{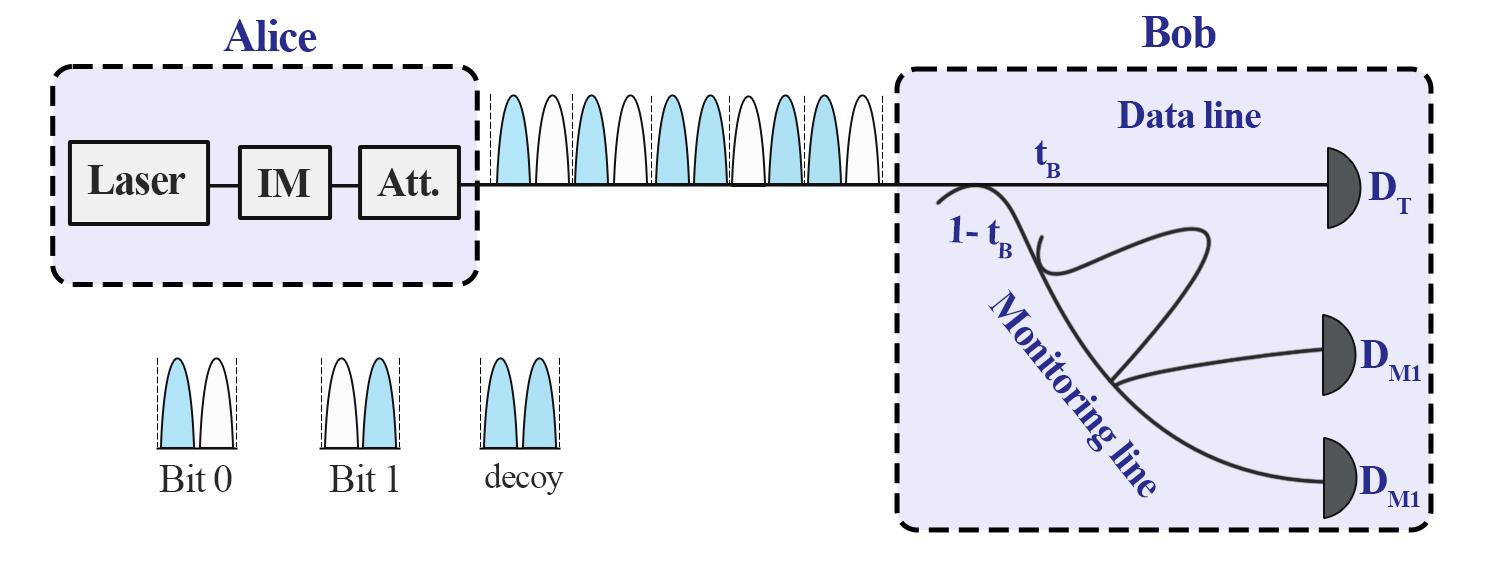}
	\caption{A schematic representation of the COW protocol. Alice prepares three types of quantum states: two signal states and one decoy state. On the receiver’s side, Bob retrieves the key from the signal states through the data line and verifies the coherence of the decoy states using the monitoring line.}
	\label{COW}
\end{figure*}

In this protocol, Alice encodes the logical bits $0$ and $1$ into the quantum states $|0\alpha\rangle$ and $|\alpha 0\rangle$, respectively. Here $|0\rangle$ denotes the vacuum state, and $|\alpha\rangle$ represents a coherent state with mean photon number $\mu = |\alpha|^2$. In addition to the signal states, Alice also transmits the decoy state $|\alpha \alpha\rangle$ to detect potential eavesdropping attempts. Each signal state is sent with probability $\tfrac{1 - f}{2}$, and the decoy state is sent with probability $f$.\\

On the receiver’s side, Bob uses an asymmetric beam splitter with transmission coefficient $t_B$ to divide the incoming pulses between the data line and the monitoring line. In the data line, the raw key is generated by measuring the arrival times of the pulse pairs using detector $D_T$. In the monitoring line, an interferometer equipped with two detectors, $D_{M1}$ and $D_{M2}$, is employed to verify the coherence between consecutive non-vacuum pulses (see Fig.~\ref{COW}). Any disturbance in the coherence measured in the monitoring line enables Alice and Bob to detect the presence of an eavesdropper.\\
\\
\section{Theoretical framework for witnessing the effective entanglement in the COW protocol}

 \label{theoretical}

In this section, we develop the theoretical framework necessary to investigate the presence of effective entanglement in the experimental data of the COW protocol. 
Before presenting the explicit form of the entanglement witnesses, it is essential to clarify the connection between entanglement-based and prepare-and-measure QKD protocols.\\

It is well known that any P\&M protocol admits a mathematically equivalent EB formulation \cite{wolf2021quantum}. In this equivalent picture, Alice and Bob share an entangled bipartite state, where a measurement on Alice’s subsystem effectively prepares the corresponding states that would be transmitted in the P\&M scheme. Moreover, Bob’s measurements in the EB formulation yields the same outcomes as in the P\&M protocol when performed on the received states.\\

Concretely, in the EB formulation, Alice measures an observable $A$ with probability $\mathbf{P}(A)$ and obtains outcome $a$ with probability $P_A(a)$ ($\sum_A \mathbf{P}(A) = \sum_a P_A(a) = 1$). This projects Bob’s subsystem onto $|\psi_{A_a}\rangle$, on which he measures $B$ to obtain outcome $b$. The resulting joint probability distribution, $P_{A,B}(a,b)$, depends on the observables $A$ and $B$ and satisfies $\sum_{a,b} P_{A,B}(a,b) = 1$ for any choice of measurement settings $A$ and $B$.
In the equivalent P\&M formulation,  Alice directly prepares the states $|\psi_{A_a}\rangle$ with probabilities $\mathbf{P}(A) P_A(a)$ and sends them to Bob, who measures $B$ and records the outcome $b$. The joint probability distributions in this case are likewise given by $P_{A,B}(a,b)$ for the measurement choices, thereby illustrating  the formal equivalence between the EB and P\&M descriptions.\\

This equivalence allows entanglement witnesses—constructed from operators $A$ and $B$—to certify the presence of effective entanglement even in P\&M implementations. More precisely, consider a P\&M protocol that produces the joint probability distributions $P_{A,B}(a,b)$. Since there exists at least one corresponding EB scenario for any P\&M protocol, there must exist an effective state $\rho_{\mathrm{eff}}$ and observables $A$ and $B$ such that
\[
\operatorname{tr}\!\big(\rho_{\mathrm{eff}} \, (\Pi_A(a) \otimes \Pi_B(b))\big) \;=\; P_{A,B}(a,b),
\]  
where  
\[
\Pi_A(a) = |\lambda_a^{(A)}\rangle \langle \lambda_a^{(A)}|, 
\qquad
\Pi_B(b) = |\lambda_b^{(B)}\rangle \langle \lambda_b^{(B)}| .
\]
Here, $|\lambda_a^{(A)}\rangle$ and $|\lambda_b^{(B)}\rangle$ are the eigenvectors of $A$ and $B$ corresponding to eigenvalues $a$ and $b$, respectively. Therefore, witnessing entanglement in $\rho_{\mathrm{eff}}$ is, by construction, equivalent to witnessing effective entanglement in the P\&M implementation.\\

The COW protocol, introduced in Section~(\ref{COWsection}), constitutes a prepare-and-measure scheme that, in principle, admits an equivalent entanglement-based formulation. The objective of the remainder of this section is to construct entanglement witnesses capable of certifying effective entanglement directly from the probability distributions observed in the prepare-and-measure realization of the COW protocol.\\

\subsection{A Family of Entanglement Witnesses for the COW Protocol}
\label{witness}

Let
\[\mathcal{H} = \mathrm{span}\{\, |n\rangle \ | \  n \in \{0,1,2,...\}\},
\]
be the Hilbert space of a single-mode field, where $|n\rangle$ denotes the Fock state containing $n$ photons. The quantum states prepared by Alice in the COW protocol therefore belong to
\[
\tilde{\mathcal{H}} = \mathcal{H} \otimes \mathcal{H}.
\]
On the other hand, let  $\mathcal{H}_A$ represent the Hilbert space of Alice's subsystem in the EB equivalent formulation of the COW protocol. The corresponding effective joint state of this formulation, denoted by $\rho_{\mathrm{eff}}$, then belongs to the set of trace-one positive operators acting on the Hilbert space $\mathcal{H}_A \otimes \tilde{\mathcal{H}}$, which we denote by $\mathcal{B}_{\mathcal{H}_A \otimes \tilde{\mathcal{H}}}$. \\

For later convenience, we introduce the following notations:
\begin{equation}
	|z+\rangle = |10\rangle, \qquad
	|z-\rangle = |01\rangle ,
\end{equation}
\begin{equation}
	|x+\rangle = \frac{|10\rangle + |01\rangle}{\sqrt{2}}, \qquad
	|x-\rangle = \frac{|10\rangle - |01\rangle}{\sqrt{2}},
\end{equation}
and define the operators $\mathbf{Z}$ and $\mathbf{X}$ acting on $\mathcal{B}_{\tilde{\mathcal{H}} }$ as
\[
\mathbf{Z} = |z+\rangle \langle z+| - |z-\rangle \langle z-|,
\ \ \ \ \ 
\mathbf{X} = |x+\rangle \langle x+| - |x-\rangle \langle x-|.
\]
Within the COW framework, $|z+\rangle$ corresponds to the presence of a single photon in the first time slot, while $|z-\rangle$ represents a single photon in the second time slot. The superposition states $|x\pm\rangle$ describe coherent delocalizations of the photon across the two time slots.\\

To witness the entanglement of $\rho_{\mathrm{eff}}$ in the COW protocol, we can, without loss of generality, set $\mathcal{H}_A=\tilde{\mathcal{H}}$ , and introduce the operator $W\in \mathcal{B}_{\tilde{\mathcal{H}} \otimes \tilde{\mathcal{H}}}$ defined as
\begin{equation}
\label{effective witness}
   W = I \otimes I \;+\; a\, \mathbf{Z} \otimes \mathbf{Z} \;+\; b\, |x+\rangle \langle x+| \otimes \mathbf{X},
\end{equation}
where $a$ and $b$ are  real parameters whose admissible ranges will be determined later. \\

In the next subsection, we will demonstrate that $W$ serves as a valid entanglement witness and identify the corresponding ranges of $a$ and $b$. Here, however, we focus on how to evaluate  the expectation value of $W$ for the effective state $\rho_{\mathrm{eff}}$  in the EB formulation,  using the probability distributions obtained from the corresponding P\&M experiment.\\

Consider first the quantity $\operatorname{tr}(\rho_{\mathrm{eff}}\, \mathbf{Z} \otimes \mathbf{Z})$, which is proportional to the second term of  $\operatorname{tr}(\rho_{\mathrm{eff}} W)$. This term can be expanded as
\begin{align}
\operatorname{tr}\!\big(\rho_{\mathrm{eff}}\, \mathbf{Z} \otimes \mathbf{Z}\big)
&= \langle z+, z+| \rho_{\mathrm{eff}} |z+, z+\rangle
+ \langle z-, z-| \rho_{\mathrm{eff}} |z-, z-\rangle \notag\\
&\quad - \langle z+, z-| \rho_{\mathrm{eff}} |z+, z-\rangle
- \langle z-, z+| \rho_{\mathrm{eff}} |z-, z+\rangle, 
\end{align}
or equivalently,
\begin{align} \label{Zexpectation}
\operatorname{tr}\!\big(\rho_{\mathrm{eff}}\, \mathbf{Z} \otimes \mathbf{Z}\big)
&= \operatorname{tr}\!\big(\rho_{\mathrm{eff}}\, \Pi_{z+}\!\otimes \Pi_{z+}\big)
+ \operatorname{tr}\!\big(\rho_{\mathrm{eff}}\, \Pi_{z-}\!\otimes \Pi_{z-}\big) \notag\\
&\quad - \operatorname{tr}\!\big(\rho_{\mathrm{eff}}\, \Pi_{z+}\!\otimes \Pi_{z-}\big)
- \operatorname{tr}\!\big(\rho_{\mathrm{eff}}\, \Pi_{z-}\!\otimes \Pi_{z+}\big),
\end{align}
where $\Pi_{z+} = |z+\rangle\langle z+|$ and $\Pi_{z-} = |z-\rangle\langle z-|$.\\

In the EB picture, the four terms appearing on the right-hand side of Eq.~(\ref{Zexpectation}) correspond to the probabilities of obtaining $|z+\rangle$ or $|z-\rangle$ outcomes after Alice and Bob perform $\mathbf{Z}$ measurements. More explicitly, $\operatorname{tr}(\rho_{\mathrm{eff}}\, \Pi_{z\circ} \otimes \Pi_{z\bullet})$, where $\circ, \bullet \in \{+, -\}$, denotes the probability that Alice obtains $|z_\circ\rangle$ and Bob finds $|z_\bullet\rangle$ as outcomes of local $\mathbf{Z}$ measurements on $\rho_{\mathrm{eff}}$. \\

Using the correspondence described at the beginning of this section, Bob’s conditional state after Alice’s measurement in the EB formulation is identical to the state that Alice would have prepared and sent in the P\&M scheme. For instance, to find the P\&M equivalent expression of $\operatorname{tr}(\rho_{\mathrm{eff}}\, \Pi_{z+} \otimes \Pi_{z+})$, one should consider the statistics of events in which Alice sends the state $|z+\rangle$ and Bob also obtains $|z+\rangle$ upon performing his $\mathbf{Z}$ measurement. The same interpretation applies to the other terms in $\operatorname{tr}(\rho_{\mathrm{eff}}\, \mathbf{Z} \otimes \mathbf{Z})$. \\

Following an analogous reasoning, one can evaluate $\operatorname{tr}(\rho_{\mathrm{eff}}\, |x+\rangle\langle x+| \otimes \mathbf{X})$, which corresponds to the next component of $\operatorname{tr}(\rho_{\mathrm{eff}} W)$. This quantity can be derived from the statistics of events where Alice sends $|x+\rangle$ and Bob identifies the received state as either $|x+\rangle$ or $|x-\rangle$ when performing his $\mathbf{X}$ measurement. \\

Summarizing, to compute the expectation value of the witness operator $W$ from the data obtained in the P\&M implementation, one requires the statistics corresponding to the cases in which Alice sends the states $|z+\rangle$, $|z-\rangle$, or $|x+\rangle$, and Bob performs measurements distinguishing between $|z+\rangle$ and $|z-\rangle$ (the $\mathbf{Z}$ basis), or between $|x+\rangle$ and $|x-\rangle$ (the $\mathbf{X}$ basis). \\

In the COW protocol, which has a P\&M structure, Bob performs a time-of-arrival measurement in the data line. This measurement perfectly distinguishes between $|z+\rangle = |10\rangle$ and $|z-\rangle = |01\rangle$, thereby implementing the desired $\mathbf{Z}$ measurement. Furthermore, it is straightforward to verify that the monitoring line distinguishes between the states $|x+\rangle = \tfrac{|10\rangle + |01\rangle}{\sqrt{2}}$ and $|x-\rangle = \tfrac{|10\rangle - |01\rangle}{\sqrt{2}}$: the former produces a detection event on detector $D_{M1}$, while the latter triggers detector $D_{M2}$. Hence, the data obtained from the monitoring line can be regarded as the outcomes of the desired $\mathbf{X}$ measurement. \\

Let us now examine the states sent by Alice. For small values of $\alpha$, one can write
\begin{align}
	\label{Alice state}
	|0\alpha\rangle   &\sim |00\rangle + \alpha |01\rangle + O(\alpha^2), \nonumber \\
	|\alpha 0\rangle  &\sim |00\rangle + \alpha |10\rangle + O(\alpha^2), \nonumber \\
	|\alpha \alpha\rangle &\sim |00\rangle + \alpha \big(|01\rangle + |10\rangle\big) + O(\alpha^2).
\end{align}
The leading-order contributions are thus the vacuum and single-photon terms. The single-photon components of the above expansions coincide precisely with the states required for evaluating the expectation value of $W$. The vacuum state $|00\rangle$ does not trigger Bob’s detectors, except through dark counts. These counts do not adversely affect the analysis as it will be discussed in the next subsection.\\

In summary, for small values of $\alpha$, and after appropriate normalization, the statistics obtained from a COW experiment can be employed to evaluate $\operatorname{tr}(\rho_{\mathrm{eff}} W)$ and thereby witness the presence of effective entanglement in the protocol.

\subsection{Conditions for a Valid Entanglement Witness}
\label{validity}

Now, it is time to discuss the reason for employing the witness operators defined in (\ref{effective witness}), and to determine the admissible region of the parameters $a$ and $b$. \\

It is well known that, for the security of any QKD protocol, the presence of effective entanglement must be verifiable solely through the measurements performed within that protocol \cite{curty2004entanglement}. As discussed in the previous subsection, in the COW protocol and for sufficiently small values of $\alpha$, Alice and Bob have access to the statistics corresponding to Alice’s effective preparation of the states $|z\pm\rangle$ or $|x+\rangle$, together with Bob’s measurements in the $X$ or $Z$ basis. Consequently, according to  \cite{curty2004entanglement}, one expects that effective entanglement can indeed be detected using these measurement statistics. \\

Among the various probabilities available to Alice and Bob in a COW setup, the quantities
\[
\operatorname{tr}\!\big(\rho_{\mathrm{eff}}\, \Pi_{z\pm}\!\otimes \Pi_{z\pm}\big)
\quad\text{and}\quad
\operatorname{tr}\!\big(\rho_{\mathrm{eff}}\, \Pi_{x+}\!\otimes \Pi_{x\pm}\big)
\]
are more relevant for identifying correlations in the protocol than
\[
\operatorname{tr}\!\big(\rho_{\mathrm{eff}}\, \Pi_{z\pm}\!\otimes \Pi_{x\pm}\big).
\]
This is because, in the COW protocol, no meaningful correlations exist between the events where Alice sends data (or decoy) pulses and Bob records clicks in the monitoring (or data) line. Consequently, terms such as $\mathbf{Z} \otimes \mathbf{X}$ do not appear in our proposed effective entanglement witness. Similarly, operators like $I \otimes \mathbf{X}$ or $\mathbf{Z} \otimes I$ correspond to local expectation values rather than genuine two-party correlations, and hence are irrelevant for entanglement detection. In fact, we have constructed the witness operator $W$ using only those probabilities that most effectively capture the correlations inherent in the COW protocol. \\

It is also worth noting that one need not be concerned about dark counts, which arise from the vacuum state $|00\rangle$. Dark counts produce equal clicks in the early and late time slots of the data-line detector, thereby reducing the correlations of interest. Hence, if the experimental data—including dark counts—still exhibit effective entanglement, we can confidently conclude that the corresponding data without dark counts would also do so. \\

We now determine the admissible region of the parameters $a$ and $b$ for which $W$ constitutes a valid entanglement witness. To this end, $W$ must not be positive semidefinite; that is, it must possess at least one negative eigenvalue. On the other hand, the expectation value of $W$ must remain nonnegative for all separable states. \\

To satisfy the first condition, it is sufficient that the smallest eigenvalue of $W$ be negative. This yields
\begin{equation}
	\lambda_{\min} = \frac{1}{2} \left( 2 - |b| - \sqrt{4a^2 + b^2} \right) < 0,
\end{equation}
which leads to the inequality
\begin{equation}
	\label{negative eigenvalue}
	1 - |b| < a^2 .
\end{equation}

To meet the second condition, it is enough to ensure that the expectation value of $W$ remains nonnegative for all pure product states, owing to the convexity of the set of separable states.  
Consider a general pure product state of the form $|\psi\phi\rangle := |\psi\rangle \otimes |\phi\rangle$, and let $r_1 = (x_1, y_1, z_1)$ and $r_2 = (x_2, y_2, z_2)$ denote the normalized Bloch vectors corresponding to $|\psi\rangle$ and $|\phi\rangle$, respectively. The expectation value of $W$ is then given by
\begin{equation}
	\langle W \rangle = \langle\psi\phi|W|\psi\phi\rangle = 1 + a\, z_1 z_2 + \frac{b}{2} (1 + x_1) x_2.
\end{equation}
We now seek the range of $a$ and $b$ for which the minimum value of $\langle W \rangle$ remains nonnegative for all $x_1, z_1, x_2$, and $z_2$ satisfying the constraints  
\[
x_1^2 + z_1^2 \le 1, \qquad x_2^2 + z_2^2 \le 1 .
\]

To simplify the analysis, we first determine the minimum value of $\langle W \rangle$ under the above constraints and subsequently examine the conditions under which it remains nonnegative.  
Define the two-dimensional vectors  
\[
\mathbf{u} = \left( \frac{b}{2}(1 + x_1),\, a z_1 \right), \qquad 
\mathbf{v} = (x_2, z_2),
\]
so that
\begin{equation}
	\langle W \rangle = 1 + \mathbf{u} \cdot \mathbf{v}.
\end{equation}
It is evident that $\langle W \rangle$ is minimized when $\mathbf{u} \cdot \mathbf{v}$ attains its minimum.  
Since $|\mathbf{v}| \le 1$, this occurs when
\[
\mathbf{v} = -\frac{\mathbf{u}}{|\mathbf{u}|},
\]
yielding $\mathbf{u} \cdot \mathbf{v} = -|\mathbf{u}|$. \\

Thus, the problem reduces to maximizing
\[
|\mathbf{u}| = \sqrt{a^2 z_1^2 + \left[ \frac{b}{2}(1 + x_1) \right]^2 },
\]
subject to the constraint  
\[
x_1^2 + z_1^2 \le 1 .
\]
Since the maximum is achieved on the boundary $x_1^2 + z_1^2 = 1$, we can parameterize the boundary by
\[
x_1 = \cos\theta, \qquad z_1 = \sin\theta,
\]
and then determine the extremum by direct differentiation. \\

After straightforward calculation, the minimum expectation value of $W$ for separable states is found to be
\begin{equation}
	\label{eq:pic3}
	\langle W \rangle_{\min} =
	\begin{cases}
		1 - |b|, & \text{if } \frac{b^{2}}{2} > a^{2}, \\[0.8em]
		1 - \dfrac{a^{2}}{\sqrt{a^{2} - b^{2}/4}}, & \text{if } \frac{b^{2}}{2} < a^{2}.
	\end{cases}
\end{equation}
Accordingly, the condition of positivity on separable states imposes the following constraints:
\begin{equation}
	\label{eq:pic2}
	\begin{cases}
		a^{2} < \dfrac{b^{2}}{2} \quad \Rightarrow \quad |b| \le 1, \\[0.5em]
		a^{2} > \dfrac{b^{2}}{2} \quad \Rightarrow \quad 4a^{4} - 4a^{2} + b^{2} \le 0.
	\end{cases}
\end{equation}

Finally, combining these constraints with the requirement that $W$ is not positive semidefinite [Eq.~(\ref{negative eigenvalue})], we identify the complete range of parameters $a$ and $b$ for which $W$ serves as a valid entanglement witness:
\begin{equation}
	\label{eq:pic1}
	\begin{cases}
		\textbf{I:} 
		\begin{cases}
			a^{2} < \dfrac{b^{2}}{2}, \\
			|b| < 1, \\
			1 - |b| < a^{2},
		\end{cases}
		\\[1.0em]
		\text{or} \\[1.0em]
		\textbf{II:} 
		\begin{cases}
			a^{2} > \dfrac{b^{2}}{2}, \\
			4a^{4} - 4a^{2} + b^{2} < 0, \\
			1 - |b| < a^{2}.
		\end{cases}
	\end{cases}
\end{equation}

\begin{figure*}[!h]
	\centering
	\includegraphics[width=0.75\linewidth]{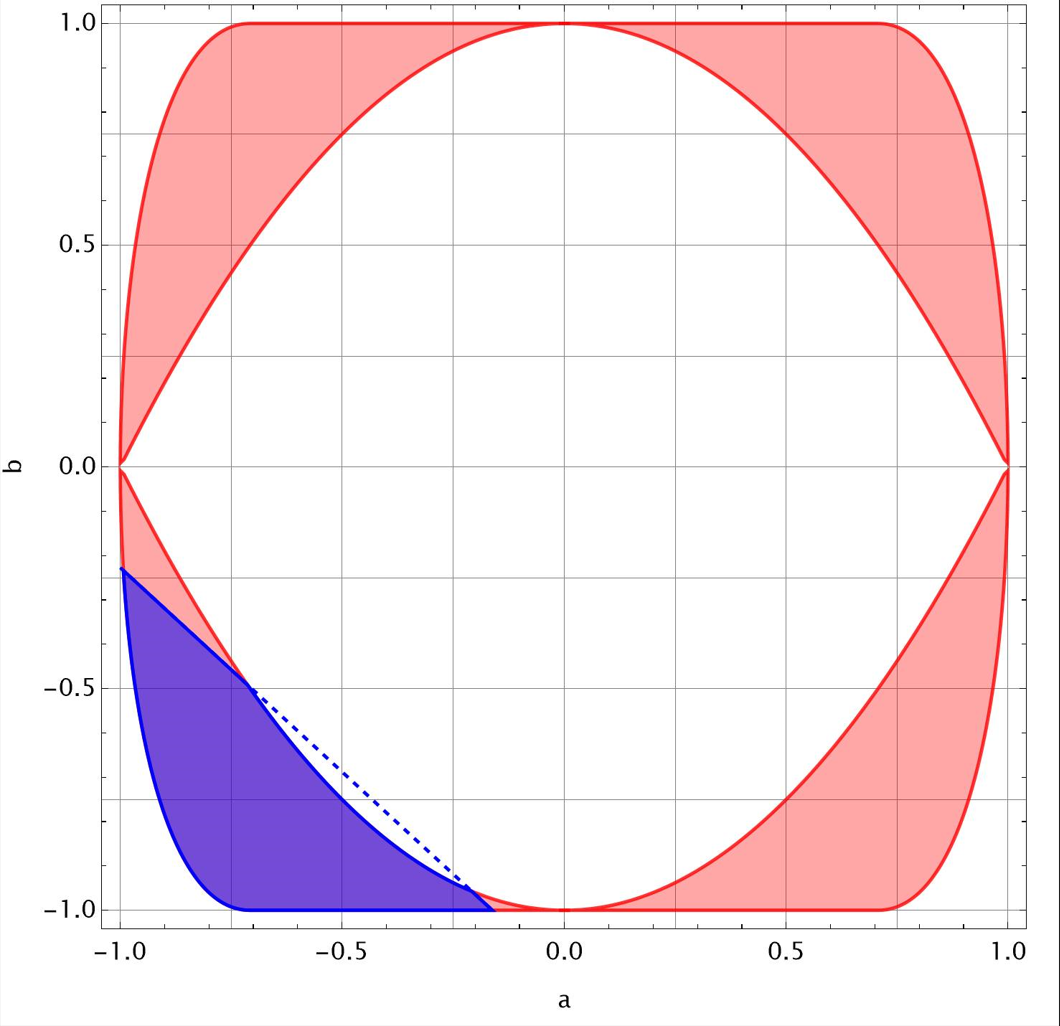}
	\caption{Schematic representation of the admissible region in the $(a,b)$ parameter space. The red area corresponds to the values of $a$ and $b$ for which $W$ qualifies as a valid entanglement witness, while the blue subset highlights the region where $W$ detects effective entanglement in the experimental data obtained under a channel loss of $14\,\mathrm{dB}$.}
	\label{acceptable}
\end{figure*}

In summary, whenever the parameters $a$ and $b$ satisfy either the conditions of set~$\mathbf{I}$ or those of set~$\mathbf{II}$, the operator $W$ qualifies as a valid entanglement witness.  
The admissible region defined by Eq.~\eqref{eq:pic1} is illustrated in Fig.~\ref{acceptable}, which clearly demonstrates that the set of operators capable of witnessing effective entanglement in the COW protocol is non-empty.

\section{Experimental Demonstration of Effective Entanglement}
\label{experimental witnessing}

After introducing a suitable family of witness operators adapted to the Coherent One-Way (COW) protocol, we now demonstrate that a subset of these operators can reveal effective entanglement within our previously reported experimental data~\cite{dadahkhani2025experimental}.\\

To this end, we first clarify how the experimentally measured quantities are utilized in evaluating the witness expectation value. Specifically, calculating $\langle W \rangle$ requires determining the expectation values of the observables $\mathbf{Z} \otimes \mathbf{Z}$ and $|x+\rangle\langle x+| \otimes \mathbf{X}$. As detailed in Section~(\ref{witness}), these quantities can be inferred from the joint statistics of Alice’s prepared optical pulses and Bob’s detection events in both the data and monitoring lines. 
For instance, evaluating $\langle \mathbf{Z} \otimes \mathbf{Z} \rangle$ involves computing the four terms defined in Eq.~(\ref{Zexpectation}), which, as discussed, correspond to the conditional probabilities of the following events: (i) Alice sends $|z+\rangle$, corresponding to the time-bin sequence $|0\alpha\rangle$ in the COW protocol, and Bob registers a click in the early time slot; (ii) Alice sends $|z+\rangle$ and Bob detects a click in the late slot; (iii) Alice sends $|z-\rangle$ (i.e., $|\alpha0\rangle$) and Bob detects a click in the early slot; and (iv) Alice sends $|z-\rangle$ and the click occurs late. 
Similarly, the expectation value of $|x+\rangle\langle x+| \otimes \mathbf{X}$ is obtained from the statistics associated with the transmission of $|x+\rangle$—that is, the two-pulse sequence $|\alpha\alpha\rangle$—and the corresponding interference outcomes recorded in the monitoring line.\\

When incorporating the experimental data into these calculations, a renormalization procedure is required to exclude events that are irrelevant to the logical bit assignments. For illustration, consider the case in which Alice transmits the sequence $|0\alpha\rangle$. In the data line, four detection outcomes are possible: (\textit{i}) a click occurs only in the early time slot; (\textit{ii}) a click occurs only in the late time slot; (\textit{iii}) clicks occur in both time slots; and (\textit{iv}) no detection event occurs. Among these, only the first two outcomes are relevant, as they correspond to Bob identifying the state sent by Alice as either $|0\alpha\rangle$ or $|\alpha0\rangle$—that is, $|z+\rangle$ or $|z-\rangle$ in the witness-operator representation. Consequently, the data are post-selected to include only these relevant events and renormalized accordingly, ensuring that whenever Alice sends $|0\alpha\rangle$, Bob’s effective outcome is restricted to identifying the state as either $|0\alpha\rangle$ or $|\alpha0\rangle$.\\

\begin{table}[h!]
\centering
\renewcommand{\arraystretch}{1.5} 
\setlength{\tabcolsep}{6.5pt}      
\caption{Renormalized probabilities obtained from the experimental implementation reported in \cite{dadahkhani2025experimental}, for channel loss of $14\,\mathrm{dB}$.
	Here $G_{S,D}$ denotes the probability that Alice sends the state $S$ and the detectors of Bob, record the $D$ click. $D_{T,e}$ and $D_{T,l}$ account for the clicks in the early and late time slots of the data line, while $D_{M1}$ and $D_{M2}$ show the two detectors of the monitoring line. The probabilities are normalized  such that $G_{\alpha 0,D_{T,e}} + G_{\alpha 0,D_{T,l}} = 1$, $G_{0\alpha,D_{T,e}} + G_{0\alpha,D_{T,l}} = 1$ and $G_{\alpha\alpha,D_{M1}} + G_{\alpha\alpha,D_{M2}} = 1$.\\
}

\begin{tabular}{|c|c|c|c|c|c|}
	\hline
	$\mathbf{G_{\alpha0,D_{T,e}}}$ & $\mathbf{G_{\alpha0,D_{T,l}}}$ & $\mathbf{G_{0\alpha,D_{T,e}}}$ & $\mathbf{G_{0\alpha,D_{T,l}}}$ & $\mathbf{G_{\alpha\alpha,D_{M1}}}$ & $\mathbf{G_{\alpha\alpha,D_{M2}}}$ \\ \hline
	\textbf{0.917124} &  \textbf{0.082876}  &  \textbf{0.0115017} & \textbf{0.884983} & \textbf{0.935484} & \textbf{0.064516} \\ \hline
\end{tabular}

\label{table}
\end{table}

The renormalized experimental data obtained from our previous implementation of the COW protocol under a channel loss of $14\,\mathrm{dB}$ are summarized in Table~\ref{table}. The presence of effective entanglement in these data can be verified using the proposed witness operator within a specific region of the parameter space $a$ and $b$. This region is depicted in Fig.~\ref{acceptable}, highlighted in blue.\\

\begin{figure*}[!h]
	\centering
	\includegraphics[width=1\linewidth]{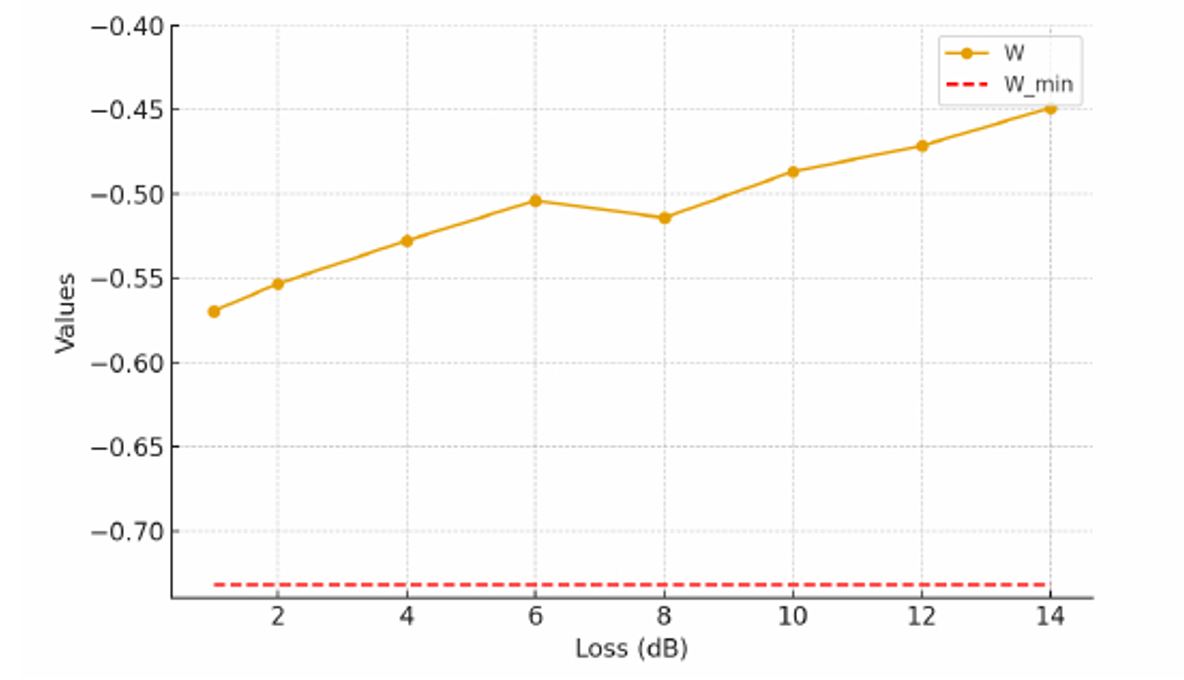}
	\caption{Expectation value of the witness operator proposed in Eq.~(\ref{effective witness}), with $a = -\tfrac{\sqrt{3}}{2}$ and $b = -\tfrac{\sqrt{3}}{2}$, for different channel losses}
	\label{different loss}
\end{figure*}

For other loss values investigated in our experiment, there also exist regions in the $(a,b)$ parameter space within the admissible domain that certify the presence of effective entanglement. Figure~\ref{different loss} shows the value of the entanglement witness as a function of channel loss for a fixed choice of $a=-\tfrac{\sqrt{3}}{2}$ and $b=-\tfrac{\sqrt{3}}{2}$. The dashed line represents the ideal case, where $tr(\rho_{\mathrm{eff}}\mathbf{Z} \otimes \mathbf{Z})=1$ and $tr(\rho_{\mathrm{eff}}|x+\rangle\langle x+| \otimes \mathbf{X})=1$. In this scenario, for example, when Alice sends $|0\alpha\rangle$, Bob’s detector would register a click in the early time slot with unit probability. 
As shown in Fig.~\ref{different loss}, increasing the channel loss causes the expectation value of the fixed witness to approach zero. This behavior is anticipated, as coherence between time bins deteriorates with increasing fiber length and, consequently, with higher channel loss. The resulting loss of coherence reduces the ability to detect effective entanglement, which is manifested as a gradual decrease in the expectation value of the witness operator.

\section{Conclusion}
\label{conclosion}

In this work, by exploiting the equivalence between prepare-and-measure (P\&M) and entanglement-based (EB) quantum key distribution protocols, we have introduced a novel method that brings entanglement‐witness techniques into the analysis of a P\&M implementation of the coherent‐one‐way (COW) protocol. Specifically, we adapted entanglement witness operators originally defined for EB schemes to be applicable to P\&M data via appropriate interpretation of experimental outcomes. This allows the direct identification of quantumness from the measurement statistics of a COW QKD implementation, which inherently follows a P\&M structure.

More concretely, we introduced a two-parameter family of entanglement witnesses, and we determined the parameter‐regions in which these operators satisfy the defining criteria of a valid witness. We applied the framework to previously obtained experimental data under different channel loss conditions (including a 14 dB loss scenario) and, after a renormalization procedure to restrict the data to relevant detection events, we computed the expectation values of these witnesses. The results reveal clear signatures of effective entanglement, demonstrably present even in realistic experimental conditions. As expected, increasing channel loss leads to a reduction in the witness expectation value, consistent with the degradation of coherence between the time-bins and the consequent reduction in quantum correlations.

By doing so, we have established a direct link between the theory of entanglement verification and practical QKD implementations.  Moreover, since prior studies have shown that the expectation value of an entanglement witness can be used to establish a lower bound on the amount of entanglement ~\cite{sun2024bounding,guhne2008lower,guhne2007estimating}, our work opens the path toward connecting experimentally observed entanglement —or effective entanglement in a P\&M context— to security guarantees in QKD. In other words, our approach suggests a route to evaluate the security of an implemented QKD protocol directly from experimentally obtained statistics, thereby bridging the gap between theoretical security proofs and laboratory practice.

In summary, our contributions are as follows: (i) we formulated a rigorous mathematical framework for verifying effective entanglement in the COW QKD protocol; (ii) we demonstrated its applicability to real experimental data, showing quantumness persists under finite‐loss conditions; (iii) we highlighted the practical impact of channel loss on entanglement witness performance; (iv) and we indicated how this opens a promising avenue for linking entanglement‐based verification directly with QKD security.  We believe this work strengthens the conceptual foundation of P\&M QKD protocols by providing a tangible tool to assess quantum correlation in practice, and we hope it will stimulate further work on deriving explicit security bounds from witness‐based metrics in prepare‐and‐measure scenarios.


\end{document}